\documentclass[9pt, sigplan, nonacm]{acmart}
\usepackage{fancyhdr} 

\usepackage{svg}

\usepackage{pifont}

\usepackage{multirow}
\usepackage{tikz}
\usepackage{pgfplots}
\usepackage{graphicx}
\usepackage[table]{xcolor}

\usepackage{etoolbox}
\usepackage{xcolor}

\usepackage{algorithmic}
\usepackage[ruled,vlined,linesnumbered]{algorithm2e}

\newcommand{\selfcite}[2][blue]{%
  \textcolor{#1}{\cite{#2}}%
}

\newcommand{\selfref}[2][blue]{%
  \textcolor{#1}{\ref{#2}}%
}

\definecolor{selfrowcolor}{RGB}{163,195,182}

\newcommand{\ie}{\textit{i}.\textit{e}.}

\def\BibTeX{{\rm B\kern-.05em{\sc i\kern-.025em b}\kern-.08em
    T\kern-.1667em\lower.7ex\hbox{E}\kern-.125emX}}
\begin{document}

\newcommand{\papertitle}{Pecker}
\newcommand{\bugterms}{bug}

\title{
    \papertitle: Bug Localization Framework for Sequential Designs via Causal Chain Reconstruction
}


\author{Jiaping Tang}
\affiliation{%
  \institution{SKLP, ICT, CAS}
  \city{Beijing}
  \country{China}
}

\author{Jianan Mu}
\affiliation{%
  \institution{SKLP, ICT, CAS}
  \city{Beijing}
  \country{China}
}

\author{Tianyun Ma}
\affiliation{%
  \institution{SKLP, ICT, CAS}
  \city{Beijing}
  \country{China}
}

\author{Zhiteng Chao}
\affiliation{%
  \institution{SKLP, ICT, CAS}
  \city{Beijing}
  \country{China}
}

\author{Jing Ye}
\affiliation{%
  \institution{SKLP, ICT, CAS}
  \city{Beijing}
  \country{China}
}

\author{Huawei Li}
\affiliation{%
  \institution{SKLP, ICT, CAS}
  \city{Beijing}
  \country{China}
}
\begin{abstract}
Debugging represents a time-consuming and labor-intensive task in hardware design, with \bugterms{} localization constituting a substantial portion of this process. While spectrum-based \bugterms{}  localization techniques have achieved remarkable success in software domains and shown promise for hardware description languages, their effectiveness severely degrades in sequential designs. Unlike software programs, hardware designs exhibit intrinsic temporal characteristics that create fundamental challenges: timing misalignment between bug activation and observation, and progressive error propagation through state elements that obscures the root cause.
To address these limitations, we propose \papertitle{}, a novel \bugterms{}  localization framework that reconstructs the broken causal chain in sequential designs. 
Our approach introduces two key innovations: temporal backtracking using Estimated Minimal Propagation Cycles to identify potential activation cycles, strategic trace pruning to eliminate state pollution effects.
We evaluate \papertitle{} on comprehensive benchmarks comprising both combinational and sequential circuits. Experimental results demonstrate that \papertitle{} effectively localizes 51\%/80\%/85\% bugs within Top-1/3/5 ranks respectively, significantly outperforming state-of-the-art techniques. 
Notably, \papertitle{} maintains robust performance across circuit complexities while existing methods exhibit severe degradation on sequential designs.

%
%
%
%

\end{abstract}



\maketitle
\section{Introduction}

Debugging is a time-consuming and labor-intensive task in hardware development~\selfcite{Dessouky_SEC_2019, Foster_2025}.
In particular, debugging sequential circuits is exceptionally challenging and costly, far more so than in combinational circuits, as designers must reason about complex temporal behaviors, state transitions, and multi-cycle fault propagation and various other timing-dependent factors.
The increasing complexity of hardware design further complicates debugging~\selfcite{Foster_DAC_2015}, with a substantial portion of the debugging process dedicated to \bugterms{} localization.
Bug localization aims to accurately identify the buggy statement responsible for the observed erroneous behavior.
The accuracy of \bugterms{} localization directly impacts the time and resources required for debugging, making the improvement of \bugterms{} localization methods critical for reducing overall debugging costs~\selfcite{Hu_TCAD_2025}.

\begin{figure}[t]
\setlength{\abovecaptionskip}{-0.00cm}
\setlength{\belowcaptionskip}{-0.00cm}
\centering
\includegraphics[width=\linewidth]{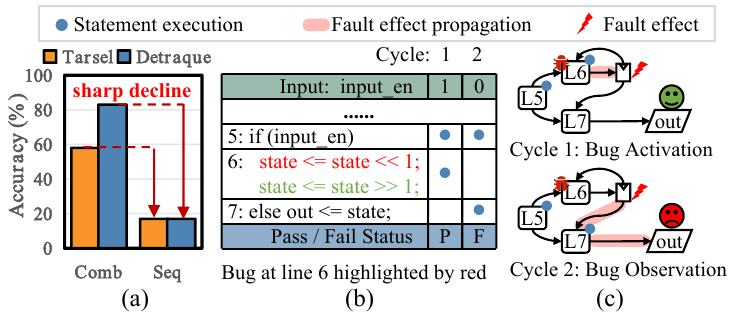}
\caption{
    (a):~Bug localization accuracy of Tarsel~\selfcite{Wu_ICCD_2022} and Detaque~\selfcite{Hu_TCAD_2025} on combinational and sequential circuits;
    (b):~A HDL code snippet along with its execution trace and statues under specified inputs, where the buggy line~(line 6) is highlighted in red, and the corresponding correct line is highlighted in green;
    (c):~A program dependency graph corresponding to (b) and the bug activates at cycle 1, propagates to the registers, and is observed at cycle 2.
}
\label{intro_data}
\end{figure}

Inspired by its success in software engineering, the hardware \bugterms{} localization community has largely adopted the Spectrum-Based Fault Localization (SBFL) paradigm~\selfcite{Wu_ICCD_2022, Wu_TCAO_2024, Hu_TCAD_2025}. 
SBFL, a major and mature research direction in software engineering~\selfcite{Wong_TSE_2016, Higor_arxiv_2017, Li_ICSE_2021, Pearson_ICSE_2017}, computes the suspiciousness of each statement by statistically analyzing its execution frequencies in passing and failing test cases. 
Statements frequently executed in passing tests are considered less suspicious, whereas those executed in failing tests receive higher scores. 
This approach benefits from simplicity, efficiency, and language generality, making it a primary technique for automated \bugterms{} localization in software languages such as Java~\selfcite{flacoco2021}. 
Its adoption for hardware description languages (HDLs) is natural, given that HDLs share many programming constructs with software, such as expressions and control-flow structures~\selfcite{Ahmad_ASPLOS_2022}.
Under the common assumption that only correct input–output (IO) pairs are available (with no access to a golden internal state), these hardware SBFL methods compare execution traces against correct IO and have shown promising performance in combinational circuits, where the bug activation and the resulting faulty output occur within the same cycle.
However, spectrum-based localization methods become notably unreliable for sequential circuits, exhibiting a pronounced degradation in localization accuracy~(shown in Figure~\selfref{intro_data}~(a)). 
The fundamental challenge arises from a timing misalignment between the cycle in which a buggy statement is activated and the cycle in which its effect becomes observable at the output—a problem unique to stateful HDL designs.

The fundamental distinction between combinational and sequential circuits lies in the presence of internal states.
In sequential circuits, outputs depend not only on current inputs but also on historically stored states. 
This difference leads to a critical debugging challenge: 
\textbf{a faulty statement may corrupt an internal state, and this erroneous state can propagate across multiple cycles before becoming visible at any observable output.}
Figure~\selfref{intro_data}~(b) illustrates this scenario with a concrete example, where a bug in line 6 (an incorrectly implemented shift direction) is activated by an input sequence. 
As shown in the execution trace, the input at cycle 1 activates the buggy statement, but the faulty output does not appear immediately.
Instead, the error propagates through register and becomes observable only at cycle 2 (Figure~\selfref{intro_data}~(c)).
This temporal misalignment between \bugterms{} activation and observation misleads spectrum-based \bugterms{} localization methods. Since the failure is detected at cycle 2, SBFL techniques incorrectly assign high suspiciousness to statements executed in that cycle—even though the actual root cause was activated earlier. 
In this example, line 7 is wrongly ranked as the most suspicious statement, while the truly buggy line 6 is overlooked.

In this paper, we propose \papertitle{}, a novel \bugterms{} localization framework that \textit{reconstructs the broken causal chain} in sequential circuits by incorporating fault propagation latency analysis and state pollution elimination.
Our approach builds on two key observations: 
\textbf{(1) distinct fault propagation latencies per statement and (2) the obfuscating effect of state pollution.}
\papertitle{} addresses these by temporally backtracking activation cycles using Estimated Minimal Propagation Cycles, strategically pruning polluted execution traces. Experimental validation demonstrates significant improvements, with \papertitle{} localizing 21/33/35 out of 41 bugs within Top-1/3/5 ranks—substantially outperforming state-of-the-art baselines while maintaining robust accuracy across sequential circuits where existing methods degrade severely.
The main contributions are summarized as follows:
\begin{itemize}
    \item We identify two fundamental challenges in sequential circuit \bugterms{} localization, \ie, the characteristic propagation latencies of faults and the obfuscating effect of state pollution on the activation-observation causal chain.
    \item We propose \papertitle{}, a novel \bugterms{} localization framework that reconstructs the broken causal chain through two key technical innovations, \ie, temporal backtracking using Estimated Minimal Propagation Cycles, strategic trace pruning to eliminate state pollution effects.
    \item We implement the proposed framework and conduct comprehensive experiments demonstrating that \papertitle{} significantly outperforms state-of-the-art baselines, locating 51\%/80\%/85\% bugs within Top-1/3/5 ranks while maintaining robust performance across circuit complexities.
\end{itemize}

\section{Background}

\begin{figure}[h]
\setlength{\abovecaptionskip}{-0.00cm}
\setlength{\belowcaptionskip}{-0.00cm}
\centering
\includegraphics[width=0.95\linewidth]{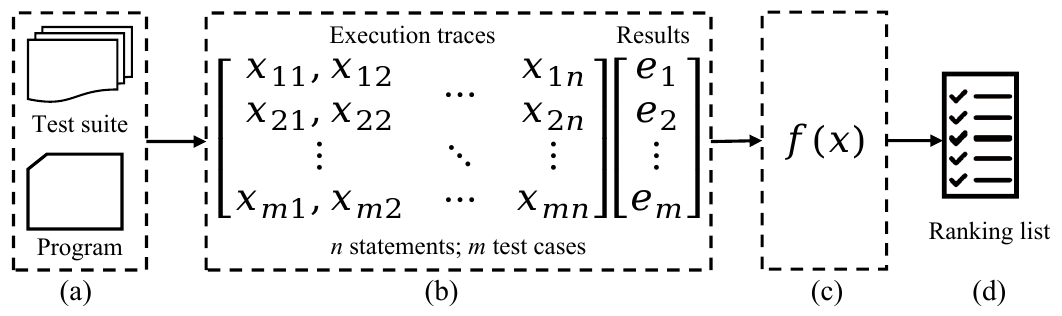}
\caption{
    The overall workflow of SBFL.
}
\label{SBFL}
\end{figure}

\subsection{Bug Localization in the Software Domain}

In the software domain, numerous effective approaches have been developed to automate bug localization~\selfcite{Wong_TSE_2016}.
Spectrum-based fault localization~(SBFL) is a more powerful and popular technique for precisely localizing bugs~\selfcite{Wong_TSE_2016, Higor_arxiv_2017, Li_ICSE_2021, Pearson_ICSE_2017}.
The workflow of SBFL is shown in Figure~\selfref{SBFL}, including data collection and suspiciousness calculation.
This method collects information about program execution, including the statement execution trace and execution results~(as shown in Figure~\selfref{SBFL}~(b)). 
By comparing execution traces from passing and failing test cases, SBFL calculates the suspiciousness of each statement with popular bug localization formula like Tarantula~\selfcite{Jones_ICSE_2002} and Ochiai~\selfcite{Abreu_2007}~(as shown in Figure~\selfref{SBFL}~(c)). The suspiciousness indicates the likelihood that it contains a bug. 
Some studies~\selfcite{Li_ICSE_2021} leverage artificial neural networks with hidden layers to learn a localization model that captures the statistical correlation between test results and statement traces. Once trained, the model is evaluated using a synthesized test dataset~(e.g., one-hot vectors) to assess the suspiciousness of each statement.


\subsection{Related Works}


Recent studies on \bugterms{} localization for hardware description languages have explored both spectrum-based~\selfcite{Wu_ICCD_2022, Wu_TCAO_2024, Hu_TCAD_2025, Hu_TODAES_2025, Heidari_TC_2025} and non-spectrum-based approaches~\selfcite{Ahmad_ASPLOS_2022, Ma_ICCAD_2025}. 
Non-spectrum-based methods, such as Cirfix~\selfcite{Ahmad_ASPLOS_2022}, employ program slicing to localize \bugterms{}s in hardware designs. Wit-HW~\selfcite{Ma_ICCAD_2025} generates witness test cases to identify buggy statements.

In contrast, spectrum-based techniques estimate statement suspiciousness using localization formula~\selfcite{Wu_ICCD_2022, Wu_TCAO_2024} or predictive models~\selfcite{Hu_TCAD_2025, Hu_TODAES_2025}.
Tarsel~\selfcite{Wu_ICCD_2022}, for instance, makes a first attempt automatically localize hardware bugs by introduced spectrum-based method that samples coverage near the error occurrence. 
Tartan~\selfcite{Wu_TCAO_2024} further improves upon Tarsel by applying minimum suspicious set data purification techniques. 
Detraque~\selfcite{Hu_TCAD_2025} explore neural network with integrating program slicing to reduce the influence of fault-irrelevant statements, and Canal~\selfcite{Hu_TODAES_2025} extends this approach by using over-sampling to balance trace information between passing and failing tests.
Although these methods improve localization accuracy, they still face challenges in handling timing misallignment bugs in sequential circuits, highlighting the need for more adaptive and precise \bugterms{} localization techniques.

\begin{figure*}[h]
\setlength{\abovecaptionskip}{-0.00cm}
\setlength{\belowcaptionskip}{-0.00cm}
\centering
\includegraphics[width=0.85\linewidth]{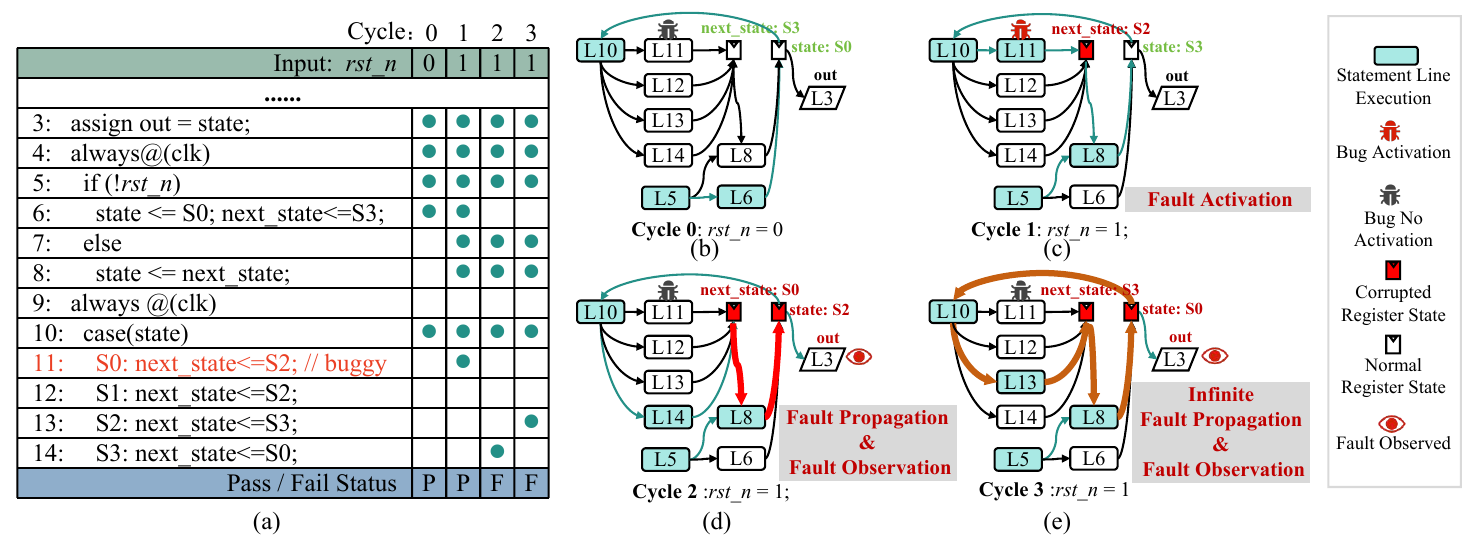}
\caption{
    Motivating Example.
}
\label{motivation_fig}
\end{figure*}

\section{Motivation and Observation}\label{motivation_section}

Locating buggy statements from erroneous outputs and execution traces is, at its core, a \textbf{causal backtracking} problem: we attempt to infer the root cause by reversing the causal chain from \emph{bug activation} to \emph{erroneous observation}. 

However, unlike combinational circuits, sequential circuits possess a large temporal state space.  
The effect of a buggy statement does not immediately appear at the output but instead unfolds as a \textbf{temporal causal chain}:
\[
\text{Activation} \;\rightarrow\; \text{Propagation} \;\rightarrow\; \text{Observation}.
\]
As a result, a bug that is activated in one cycle typically does \emph{not} produce an observable error in the same cycle.

Therefore, the \textbf{central idea of this work} is to design a localization method that explicitly models this \textbf{temporal causal chain}.

Using the example in Figure~\selfref{motivation_fig}, we illustrate this chain and derive two key observations that motivate our approach.  
Figure~\selfref{motivation_fig}~(a) shows a snippet containing a bug at line~11: the correct assignment should be \texttt{next\_state <= S1}, but it is incorrectly written as \texttt{next\_state <= S2}.  
During execution, the input at cycle~1 activates this buggy line, and the resulting error becomes observable at cycle~2.  
Figures~\selfref{motivation_fig}~(b)--\selfref{motivation_fig}~(d) present the program dependence graph (PDG) for each cycle, showing how the activation at cycle~1 propagates through intermediate states before reaching the output.
As shown in Figure~\selfref{motivation_fig}~(b), the circuit completes its initialization at cycle 0. 
Immediately afterward, the buggy statement is executed, activating the \bugterms{} and storing the corrupted value in the internal register \textit{next\_state}, shown in Figure~\selfref{motivation_fig}~(c). 
At cycle 2, executing line 8 further propagates the corrupted value into the internal register \textit{state}. Since \textit{state} directly drives the output, the fault becomes observable at this point, shown in Figure~\selfref{motivation_fig}~(d).
Because both next-state and state have already been corrupted, the subsequent executions will continue to produce observable faulty outputs~(shown in Figure~\selfref{motivation_fig}~(e)), even if the buggy statement is not executed.

From this example, we derive two motivating observations for sequential circuit bug localization:

\textbf{Observation 1: Faults Exhibit Characteristic Propagation Latencies.}
Our first observation is that different statements exhibit distinct latencies when propagating errors to outputs.
Each statement has a \textit{Estimated Minimal Propagation Cycle (EMPC)}, defined as the minimal number of cycles required for a fault originating from that statement to become observable at any primary output.
As the PDGs in Figure~\selfref{motivation_fig}~(c) and (d) illustrate, fault effects on different statements~(assumed at lines 11 and 8) have different temporal propagation delays between activation and the output node.  
Thus, the mapping from input stimuli to output observations alone is insufficient for accurate backtracking.  
To reliably locate bugs in a large sequential state space, we must perform \textbf{cross-cycle reasoning} that reflects each statement's propagation characteristics.



\textbf{Observation 2: State Pollution Obfuscates the Causal Trail.}
Once a state is corrupted, that error contaminates all computations that depend on it in subsequent cycles, a phenomenon we refer to as \textit{state pollution}.
This means an observed failure might not be the direct result of the statements executed in the current cycle, but rather a lingering effect of a \bugterms{} activated many cycles earlier.
As shown in Figure~\selfref{motivation_fig}~(d), the bug at line~11 produces an incorrect register state that will cause infinite erroneous outputs in later cycles, as shown in Figure~\selfref{motivation_fig}~(e).
Such pollution introduces ambiguity: when an error is observed in a later cycle, it is unclear whether it is due to a \emph{new} activation or the continued influence of a \emph{previous} corrupted state.
To mitigate this ambiguity, we prune all observations after the first erroneous internal state.  
This aligns with common engineering practice for debugging sequential circuits: the first incorrect internal state or output  is typically the most informative.  

\section{\papertitle~Framework}\label{sec:framework}
\begin{figure*}[htbp]
\setlength{\abovecaptionskip}{-0.00cm}
\setlength{\belowcaptionskip}{-0.00cm}
\centering
\includegraphics[width=\linewidth]{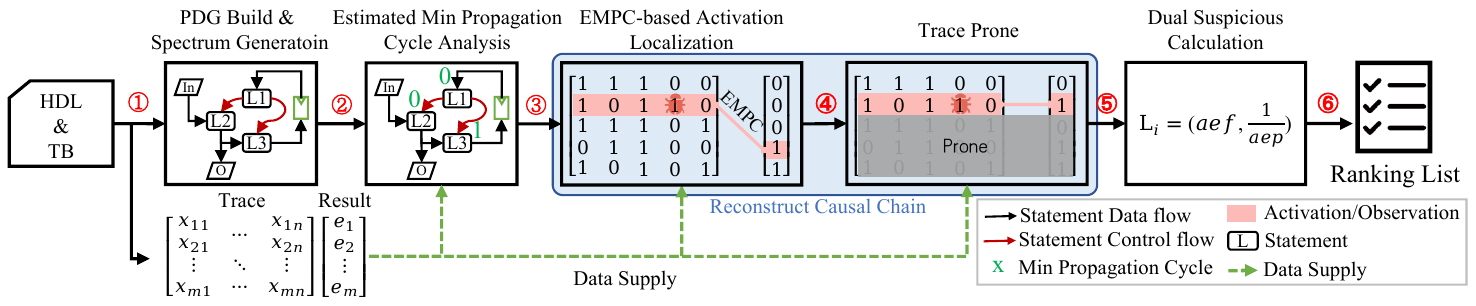}
\caption{
    The framework of \papertitle.
}
\label{framework}
\end{figure*}
The \papertitle{} framework aims to address the \textit{temporal uncertainty} problem by reconstructing the causal chain from \bugterms{} activation to observation. 
As illustrated in Figure~\selfref{framework}, our approach takes the HDL design and its testbench as inputs and produces a ranked list of suspicious statements.
The framework operationalizes our two key observations through a cohesive workflow: 
it first pinpoints the potential origin of errors in time by leveraging unique fault propagation latencies, then isolates the initial fault effect by strategically trimming polluted execution traces. 
This process enables a dual-suspiciousness scoring scheme that effectively ranks the statements, which consists of the following steps.


\subsection{Building the Program Dependency Graph for Hardware Designs}
\label{ssec:PDG}
To enable cross-cycle fault analysis, we adapt the Program Dependency Graph (PDG) from software engineering~\selfcite{Ferrante_pdg} to the context of hardware description languages to  captures dependencies within hardware designs. 
The primary adaptation involves extending the traditional PDG to explicitly model the temporal data flow facilitated by registers.
Our PDG construction algorithm (Algorithm~\selfref{alg_sdg}) proceeds in two phases:

\begin{itemize}
    \item \textbf{Control Dependency Construction (Lines 3-5):} We first establish control dependencies by traversing the AST, following conventional methods.
    \item \textbf{Data Dependency Augmentation (Lines 6-10):} We then augment the graph with data dependencies critical for hardware. A key aspect is treating registers as explicit nodes. Data dependency edges connect statements that write to a register with the corresponding register node, and from that register node to all statements that read its value in subsequent cycles. This captures the propagation of values across clock boundaries.
\end{itemize}
The resulting PDG serves as the structural foundation for our subsequent propagation latency analysis, modeling how fault effects can traverse from any statement to observable outputs over multiple cycles.


\begin{algorithm}[t]
\caption{Program Dependency Graph}
\label{alg_sdg}
\KwIn{HDL design code}
\KwOut{Program dependency graph of the HDL}

\SetKwFunction{FExtractControl}{ExtractControlFlow}
\SetKwFunction{FExtractData}{ExtractDataFlow}

AST $\gets$ parse HDL into an abstract syntax tree\;
PDG $\gets$ an empty graph\;

\SetKwProg{Fn}{Function}{}{}

\For{each node in AST}{
    \If{node is a control statement}{
        PDG.addControlEdge(node.parent, node)
    }
}

\For{each node in PDG}{
    lhs $\gets$ left-hand side of node.statement\;
    rhs $\gets$ right-hand side of node.statement\;
    PDG.addDataEdge(node, lhs.node)\;
    PDG.addDataEdge(rhs.node, node)\;
}
\Return PDG



\end{algorithm}

\subsection{Analyzing Estimated Minimal Propagation Cycles}
\label{ssec:empc}
With the PDG in place, we tackle the first core challenge: estimating \textit{when} a \bugterms{} might have been activated. 
Our approach is based on the observation that bugs located in different statements exhibits varying propagation delays before their effects are observed at the output ports. 
We define the \textbf{Estimated Minimal Propagation Cycle (EMPC)} of a statement as the estimated minimum number of clock cycles required for a fault effect originating from it to reach any primary output.

Algorithm~\selfref{alg_prop} details the EMPC computation as follows:
\begin{itemize}
    \item \textbf{Initialization (Lines 1-4):} The algorithm initializes by setting the EMPC of all observable output nodes to 0, as a fault effect is immediately visible there. 
    Register nodes are assigned an intrinsic delay of 1 cycle, representing the clock boundary crossing. Combinational nodes have a delay of 0.
    \item \textbf{Backward Propagation from First Failure (Lines 5-7):} The analysis starts from the first cycle where a failure is observed ($C_{obs}$). We perform a backward traversal on the subgraph of the PDG that is \textit{activated by the execution trace}. 
    \item \textbf{Iterative EMPC Update (Lines 9-18):} For each node encountered in the \texttt{DynamicProp} function, we update the EMPC values of its predecessor nodes if a shorter propagation cycle from it to an output is found. The update follows:
    \begin{equation}
        EMPC_{new} = EMPC_{node} + Delay_{node}
    \end{equation}
    This process repeats across the trace, moving cycle-by-cycle backwards from ($C_{obs}$), until the EMPC values for all nodes in the trace subgraph converge to their estimated minimums.
\end{itemize}

This step provides an estimated answer to the question: "If a fault was observed at  ($C_{obs}$), what is the latest cycle it could have been activated?" We use the estimated \textit{minimum} propagation cycle because the maximum can be unbounded in designs with sequential loops, while the estimated minimum provides a practical bound for causal reasoning.

\begin{algorithm}[h]
\caption{Estimated Minmial Propagation Cycle}
\label{alg_prop}
\KwIn{PDG, Traces, Results}
\KwOut{Min propagation cycle for each statement}

\SetKwProg{Fn}{Function}{}{}

stmt\_2\_prop $\gets$ initial $\infty$ for all nodes in PDG\;
stmt\_2\_prop[node] = 0 $\gets$ for all output nodes\;

node.dealy = 0 $\gets$ for all nodes in PDG\;
node.dealy = 1 $\gets$ for all regs in PDG\;
\SetKwFunction{FDynamicProp}{DynamicProp}
fail\_cycle $\gets$ find the first fail cycle in Results\; 
\For{cycle\_idx from fail\_cycle to 0}{
    \FDynamicProp{Traces[cycle\_idx]}\;
}
\Return stmt\_2\_prop

\Fn{\FDynamicProp{trace}}{
    queue $\gets$ outputs\;
    \While{queue not empty}{
        head = queue.pop\_front()\;
        \For{pred in predecessors(head)}{
            \If{trace[pred]}{
                delay = head.dealy + stmt\_2\_prop[head]\;
                \If{delay < stmt\_2\_prop[pred]}{
                    queue.add(pred)\;
                    stmt\_2\_prop[pred] = delay\;
                }
            }
            
        }
    }
}
\end{algorithm}

\subsection{Localizing Potential Activation Cycles}
\label{ssec:activation}
The EMPC value provides a temporal lever to connect an observed failure to its potential root cause. For
each statement $S$ in the execution trace, we localize its potential activation cycle $C_{act}(S)$ by working backward from the first observed failure cycle $C_{obs}$:

\begin{equation}
    C_{act}(S) = C_{obs} - EMPC(S)
\end{equation}

This equation embodies our core causal reasoning: \textbf{If statement $S$ is the true buggy statement, and its fault effect was first observed at $C_{obs}$, then it must have been activated at or before cycle $C_{act}(S)$.} This step transforms the \bugterms{} localization problem from searching the entire execution trace for each statement to validating a single, high-probability candidate cycle.


\subsection{Pruning Execution Traces to Isolate Initial Fault Effect}
\label{ssec:pruning}
To address the challenge of state pollution, where a single activated fault causes cascading failures that corrupt subsequent states, we present a trace pruning step. This step eliminates the noise from these secondary effects, ensuring our analysis focuses solely on the initial fault manifestation.

Once the potential activation cycle $C_{act}(S)$ for a statement $S$ is identified, we prune the execution trace for $S$ by discarding all cycles following $C_{act}(S)$. This operation produces two key benefits: it eliminates the noise introduced by state pollution, as any correlation between statement execution and test failure becomes unreliable after the internal state has been corrupted; and it creates a clean context for scoring, where $C_{act}(S)$ serves as the definitive point of fault manifestation while all preceding cycles provide the passing execution context.
This step is applied individually for each statement candidate, creating a tailored, pollution-free context for evaluation.


\subsection{Calculating Dual Suspiciousness Scores}
\label{ssec:scoring}

To effectively rank statements using the cleaned traces from the previous step, we introduce a dual-score scheme. Conventional SBFL formulas relying on a single score prove less effective in our context, where execution traces are predominantly composed of passing cycles after pruning.

Our proposed scoring mechanism utilizes two complementary metrics:
\begin{equation}
    \text{Suspiciousness} = (aef, \frac{1}{aep})
\end{equation}

The first component, \textbf{Activation-Execution-Failure (aef)}, represents the count of times a statement was executed in its activation cycle. The second component, \textbf{Inverse Activation-Execution-Pass ($\frac{1}{aep}$)}, is the reciprocal of the count of times a statement was executed before the activation cycle.

Statements are initially filtered to include only those with non-zero $aef$ values, ensuring they are implicated in at least one failure. The final ranking is determined by sorting candidates primarily by $aef$ in descending order, and secondarily by $\frac{1}{aep}$ in descending order. This approach effectively prioritizes statements that are consistently present when failures occur while being infrequently involved in passing executions.

\section{Experiment}


\subsection{Experimental Setup}

\subsubsection{Implementation}
\papertitle{} is implemented in Python 3.8.
The framework leverages Pyverilog~\selfcite{Pyverilog_ASPDAC_2015} to parse HDLs and construct the abstract syntax tree.
Simulation traces are collected using Synopsys VCS-\textit{2016}.
All experiments are conducted on a server equipped with Intel Xeon Platinum CPUs under Ubuntu 20.04.
\subsubsection{Evaluation Metrics}
To quantitatively evaluate the effectiveness of \bugterms{} localization, we use the following widely used measurements.
\begin{itemize}
    \item \textit{Top-K}~\selfcite{Kochhar_2016_topk}: Top-K represents the number of successfully localized bugs within Top-K positions. We consider K values of 1, 3, and 5, as recommended in previous studies~\selfcite{Wu_ICCD_2022, Hu_TCAD_2025}. A higher Top-K value indicates a greater accuracy in localizing \bugterms{}s.
    \item \textit{Mean First Rank~(MFR)}~\selfcite{Li_2017_MFR}: MFR calculates the mean rank of the buggy statements. A lower-MFR value indicates better accuracy. The lower bound of MAR value is 1, which means all bugs are localized at Top-1.
\end{itemize}

\subsubsection{Benchmarks}
As shown in Table~\selfref{bechmark}, in order to evaluate \papertitle, we use the benchmark based on Tarsel~\selfcite{Wu_ICCD_2022} and Wit-HW\selfcite{Ma_ICCAD_2025}, which consists of 12 designs and 41 bugs. Each bug is paired a bug-triggering test case written by verification engineers.
Furthermore, the benchmarks are categoried into three levels of difficulty. 
The easy level includes small combinational designs. 
The medium level features small sequential designs. 
The hard level contains some industrial designs.

\begin{table}[h]
\centering
\setlength{\abovecaptionskip}{-0.00cm}
\setlength{\belowcaptionskip}{-0.00cm}
\caption{Hardware bug localization benchmarks.}
\label{bechmark}
\resizebox{0.8\linewidth}{!}{
\begin{tabular}{
        c|c|c|c
    }
\hline
\textbf{Category} & \textbf{Design} & \textbf{Size~(LOC)} & \textbf{\#~Bugs} \\
\hline
\hline
\multirow{2}{*}{Easy} 
    & decoder                & 25 & 6 \\
    \cline{2-4}
    & alu                    & 37 & 6 \\
\hline
\multirow{6}{*}{Medium} 
    & counter                & 56 & 2 \\
    \cline{2-4}
    & lshift                 & 30 & 3 \\
    \cline{2-4}
    & led\_controller        & 76 & 4 \\
    \cline{2-4}
    & aribiter               & 112 & 3 \\
    \cline{2-4}
    & fsm\_full              & 115 & 2 \\
    \cline{2-4}
    & fsm\_16                & 132 & 4 \\
\hline
\multirow{4}{*}{Hard} 
    & sdram\_controller      & 420 & 3 \\
    \cline{2-4}
    & sha3                   & 499 & 3 \\
    \cline{2-4}
    & i2c                    & 1233 & 3 \\
    \cline{2-4}
    & reed\_decoder          & 4366 & 2 \\
\hline
\end{tabular}
}
\end{table}

\subsubsection{Baselines}
To compare \papertitle{} with various \bugterms{} localization tools, we choose several representative works:
\begin{itemize}
    \item Tarsel~\selfcite{Wu_ICCD_2022}: The first work to introduce spectrum-based \bugterms{} localization techniques to hardware description languages.
    \item Detraque~\selfcite{Hu_TCAD_2025}: Uses a learning-based method to model and analyze spectrum data for \bugterms{} localization.
    \item Wit-HW~\selfcite{Ma_ICCAD_2025}: Strengthens \bugterms{} localization by generating witness test cases.
\end{itemize}

\begin{table*}[h]
\centering
\setlength{\abovecaptionskip}{-0.0cm}
\setlength{\belowcaptionskip}{-0.0cm}
\caption{Hardware bug localization effectiveness comparison~(Comb: Combinational, Seq: Sequential).}
\label{overall_cmp}
\resizebox{0.8\linewidth}{!}{
\begin{tabular}{
        c|c|c|c c|c c|c c | c c
    }
\midrule
\textbf{Design Type} & \textbf{Category} & \textbf{Approach} & 
    \textbf{Top-1} & 
    $\%_{\textbf{Top-1}}$ & \textbf{Top-3} & $\%_{\textbf{Top-3}}$ & \textbf{Top-5} & $\%_{\textbf{Top-5}}$ & \textbf{MAR} & $\Uparrow_{MAR}$ \\
\midrule
\midrule
\multirow{4}{*}{\rotatebox{0}{\textbf{Comb}}} & 
\multirow{4}{*}{Easy} 
   & Tarsel      & 7 & 58\% & 10 & 83\% & 11 & 92\% & 2.1 & 1.6$\times$ \\
   &  & Detraque    & 10 & 83\% & 12 & 100\% & 12 & 100\% & 1.5 & $1.2\times$ \\
   &  & Wit-HW      & 8 & 67\% & 9 & 75\% & 9 & 75\% & 2.67 & $2.1\times$ \\
    \rowcolor{selfrowcolor}
   \multirow{0}{*}{\cellcolor{white}} & \multirow{0}{*}{\cellcolor{white}} 
   &  \papertitle & 10 & 83\% & 12 & 100\% & 12 & 100\% & 1.3 & $1.0\times$ \\
\midrule
\multirow{8}{*}{} & 
\multirow{4}{*}{Medium} 
    &  Tarsel      & 3 & 17\% & 6 & 33\% & 10 & 56\% & 12.3 & 2.0$\times$ \\
    & & Detraque    & 2 & 11\% & 9 & 50\% & 11 & 61\% & 12.2 & 1.9$\times$ \\
    & & Wit-HW      & 6 & 33\% & 7 & 39\% & 8 & 44\%  & 13.9 & 2.2$\times$ \\
    \rowcolor{selfrowcolor}
    \cellcolor{white} \textbf{Seq} & \cellcolor{white} 
    &  \papertitle & 7 & 39\% & 16 & 89\% & 17 & 94\% & 6.3 & 1.0$\times$ \\
\cline{2-11}
    &
\multirow{4}{*}{Hard} 
    & Tarsel      & 1 & 9\% & 1 & 9\% & 3 & 27\% & 23.1 & 1.1$\times$ \\
    & & Detraque    & 1 & 9\% & 3 & 27\% & 4 & 36\% & 26.6 & 1.2$\times$ \\
    & & Wit-HW      & 2 & 18\% & 2 & 18\% & 3 & 27\% & 29.6 & 1.4$\times$ \\
    \rowcolor{selfrowcolor}
    \multirow{0}{*}{\cellcolor{white}} & \multirow{0}{*}{\cellcolor{white}} 
    &  \papertitle & 4 & 36\% & 5 & 45\% & 6 & 55\% & 21.8 & 1.0$\times$ \\
\midrule
\multirow{4}{*}{\textbf{Overall}}
    & - & Tarsel      & 11 & 27\% & 17 & 41\% & 24 & 59\% & 12.2 & 1.4$\times$ \\
    & - & Detraque    & 13 & 32\% & 24 & 59\% & 27 & 66\% & 12.9 & 1.4$\times$ \\
    & - & Wit-HW      & 16 & 39\% & 18 & 44\% & 20 & 49\% & 14.8 &  1.7$\times$ \\
    \rowcolor{selfrowcolor}
    \multirow{0}{*}{\cellcolor{white}} & \cellcolor{white}-
    &  \papertitle & 21 & 51\% & 33 & 80\% & 35 & 85\% & 9.0 & 1.0$\times$ \\

\midrule
\end{tabular}
}
\end{table*}

\subsection{Overall Effectiveness}
\label{ssec:overall_effectiveness}
Table~\selfref{overall_cmp} presents the overall effectiveness of various \bugterms{} localization tools.
Across both Top-K and MFR metrics, \papertitle{} consistently achieves the highest \bugterms{} localization accuracy for both combinational and sequential circuits.
Specifically, \papertitle{} successfully localizes 21, 33, and 35 bugs within the Top-1, Top-3, and Top-5 ranks, respectively, substantially outperforming Tarsel (11/17/24), Detraque (13/24/27), and Wit-HW (16/18/20). 
In terms of MFR, \papertitle{} achieves a value of 9.0, which is notably lower than Tarsel (12.2), Detraque (12.9), and Wit-HW (14.8), further confirming its effectiveness.


We further analyze the effectiveness of different tools across bugs of varying difficulty levels. 
For the easy category, consisting of combinational circuits, all tools except Wit-HW achieve high \bugterms{} localization accuracy, with over 90\% of bugs ranked within the Top-5.
This high performance stems from the nature of combinational circuits, where \bugterms{} activation and observation occur within the same cycle.
As a result, observing a fault guarantees that the corresponding buggy statement has been executed, providing a direct correlation between statement execution and output errors.

In contrast, for the medium category comprising sequential circuits, all tools except \papertitle{} exhibit significant degradation in localization performance.
For instance, in terms of Top-5 accuracy, Tarsel drops from 92\% to 56\%, Wit-HW from 75\% to 44\%, while \papertitle{} experiences only a marginal decrease of 6\%. 
This disparity arises because existing tools fail to account for the temporal causal chain between \bugterms{} activation and observation inherent in sequential circuits. 
\papertitle{} explicitly addresses this issue by identifying the \bugterms{} activation cycle and pruning subsequent noisy execution traces, which constitutes a key contribution of this work.

For the hard category, all tools exhibit a decline in localization accuracy; nevertheless, \papertitle{} still maintains the highest accuracy among all compared tools, demonstrating its robustness across circuits of varying complexity.
In summary, \papertitle{} demonstrates the highest \bugterms{} localization accuracy for both combinational and sequential circuits.

\subsection{Ablation Study}
\subsubsection{Analysis of Component Effects}
To evaluate the contribution of activaton localization and noisy trace pruning, we implement two ablation variants: one removing activation localization~(\papertitle{} w/o AL) and another removing noisy trace pruning~(\papertitle{} w/o NTP).
As shown in Table~\selfref{ablation}, removing either activation localization or noisy trace pruning leads to a noticeable drop in localization accuracy.
For example, under the Top-1 metric, removing activation localization or noisy trace pruning results in an 8\% and 19\% drop in accuracy, respectively.
These results demonstrate that both components play essential roles in the overall effectiveness of the approach.
\vspace{-5pt}
\begin{table}[h]
\centering
\setlength{\abovecaptionskip}{-0.00cm}
\setlength{\belowcaptionskip}{-0.00cm}
\caption{Ablation study of different components.}
\label{ablation}
\resizebox{\linewidth}{!}{
\begin{tabular}{
        c|c c|c c|c c 
    }
\midrule
\textbf{Approach} & 
\textbf{Top-1} & 
$\%_{\text{Top-1}}$ & 
\textbf{Top-3} & 
$\%_{\text{Top-3}}$ & 
\textbf{Top-5} & 
$\%_{\text{Top-5}}$ \\
\midrule
\midrule

\papertitle & 21 & 51\% & 33 & 80\% & 35 & 85\%  \\

\papertitle{} w/o AL & 18 & 43\% & 27 & 65\% & 28 & 68\% \\

\papertitle{} w/o NTP & 13 & 32\% & 22 & 53\% & 28 & 68\% \\

\midrule
\end{tabular}
}
\end{table}

\subsubsection{Analysis of EMPC}
\begin{figure}[h]
\setlength{\abovecaptionskip}{-0.00cm}
\setlength{\belowcaptionskip}{-0.00cm}
\centering
\includegraphics[width=0.7\linewidth]{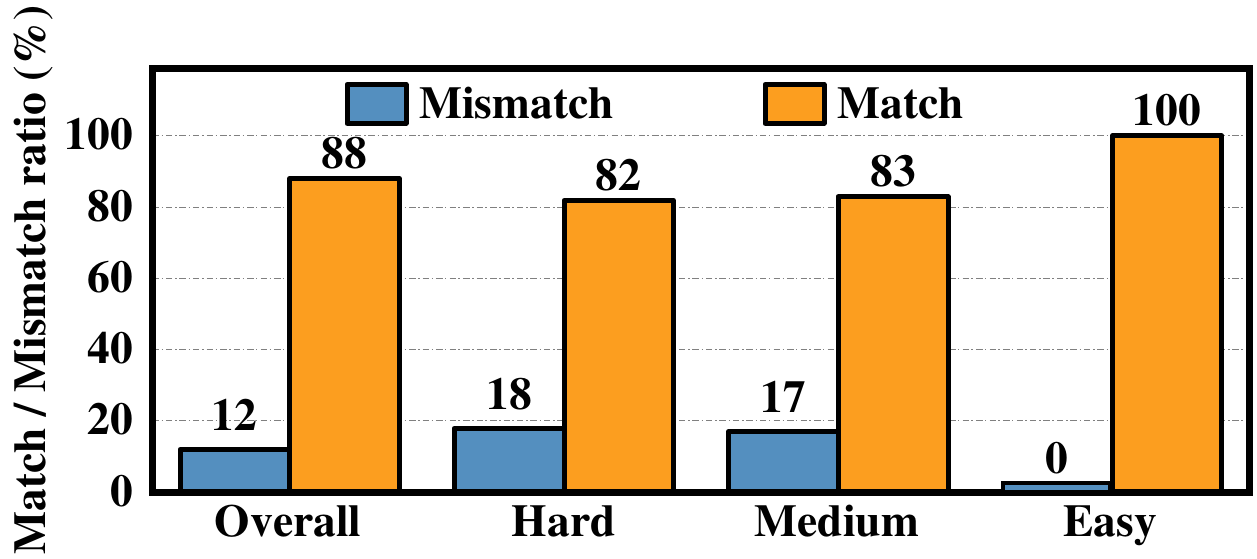}
\caption{
    Match and mismatch ratio between estimated and true activation cycles.
}
\label{mpc_cmp}
\end{figure}

To further investigate why the estimated minimum propagation cycle is effective for identifying the bug activation cycle, we compare the estimated activation cycle calculated by EMPC with the ground-truth activation cycles of buggy statements across all benchmarks. 
As shown in Figure~\selfref{mpc_cmp}, we present the matching accuracy between the estimated and true activation cycles across circuits of varying difficulty levels.
Overall, the estimated activation cycles align well with the ground-truth activation cycles, achieving an average match ratio of 88\%.
This strong correspondence indicates that EMPC provides a reliable approximation of the actual fault propagation latency, thereby enabling more precise localization of the root-cause statements and further validating the effectiveness of our approach.

\subsubsection{Analysis of Trace}
\setlength{\abovecaptionskip}{-0.00cm}
\setlength{\belowcaptionskip}{-0.00cm}
\begin{figure}[h]
\centering
\includegraphics[width=0.7\linewidth]{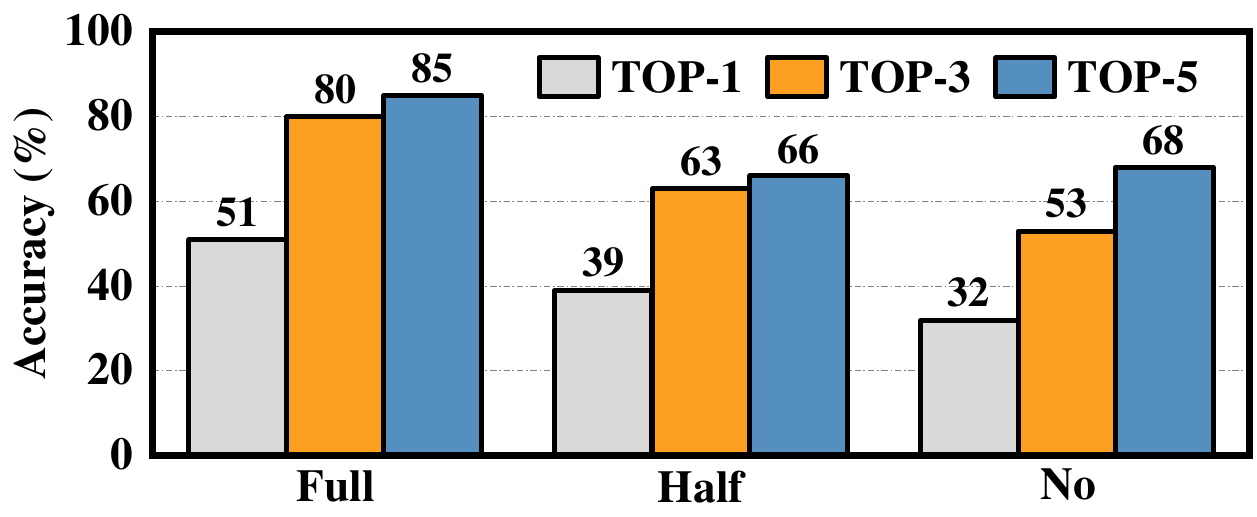}
\caption{
    Bug localization accuracy under Full, Half and No trace truncation.
}
\label{trace_cmp}
\end{figure}

To further examine the impact of noisy traces on localization accuracy, we design three levels of trace truncation: full truncation, half truncation, and no truncation.
Full truncation yields a noise-free trace, while no truncation produces the highest noise level.
As shown in Figure~\selfref{trace_cmp}, full truncation consistently delivers the highest Top-1/3/5 accuracy, validating the effectiveness of our noise-pruning strategy.
However, we also observe that for the Top-5 metric, no truncation yields only slightly better accuracy than half truncation. 
A plausible explanation is that a small portion of the later trace may still contain non-noisy information that contributes to \bugterms{} localization.
In future work, we plan to investigate how to selectively extract such informative segments from noisy traces to further enhance localization accuracy.

\section{Conclusion}

This paper proposes \papertitle{}, an automated \bugterms{} localization framework for HDLs that reconstructs the causal relationship between \bugterms{} activation and observation, addressing the misalignment between activation and observation in sequential circuits. 
Specifically, we introduce an EMPC-based method to locate the \bugterms{} activation cycle.
Besides, we propose execution trace truncation to remove the non-causal effects introduced by corrupted states.
Experimental results demonstrate that \papertitle{} effectively localizes 51\%/80\%/85\% bugs within Top-1/3/5 ranks respectively, significantly outperforming state-of-the-art techniques, especially for sequential circuits.



\bibliographystyle{ACM-Reference-Format}
\bibliography{references}

\end{document}